\documentclass{elsart}

\def\g21{G21.5-0.9}
\def\chan{$CHANDRA$}
\def\xmm{$XMM$}
\def\pshock{$pshock$}

\usepackage{natbib}
\usepackage{graphicx}
\usepackage{amssymb}

\begin{document}
\begin{frontmatter}
\title{The Plerionic Supernova Remnant \g21: In and Out}
\author{Heather Matheson}
\author{and Samar Safi-Harb}
\address{Department of Physics and Astronomy, University of Manitoba,
  Winnipeg, Manitoba, R3T 2N2, CANADA matheson@physics.umanitoba.ca,
  samar@physics.umanitoba.ca}
\centerline{\it{\small Advances in Space Research, in press}}

\begin{abstract}
The absence of a supernova remnant (SNR) shell surrounding the Crab and other
plerions (pulsar wind nebulae) has been a mystery for 3 decades.
\g21\ is a particularly intriguing plerionic
SNR in which the central powering engine is not yet detected.  Early
\chan\ observations revealed a faint extended X-ray halo which
was suggested to be associated with the SNR shell; however its
spectrum was nonthermal, unlike what is expected from an SNR shell.
On the other hand, a plerionic origin to the halo is problematic since
the X-ray plerion would be larger than the radio plerion.
We present here our analysis of an integrated 245~ksec of archival \chan\ data
acquired with the High-Resolution Camera (HRC) and 520~ksec acquired with
the Advanced CCD Imaging Spectrometer (ACIS).  This study provides the
deepest and highest resolution images obtained to date. The
resulting images reveal for the first time 1) a limb-brightened
morphology in the eastern section of the halo, and 2) a rich structure in the
inner (40$^{\prime\prime}$-radius) bright plerion including wisps and a
double-lobed morphology with an axis of symmetry running in the
northwest-southeast direction. Our spatially resolved spectroscopic study
of the ACIS-I data indicates that 
the photon index steepens with increasing distance from the central point 
source out to a radius of 40$^{\prime\prime}$ then becomes constant at 
$\sim$2.4 in the X-ray halo (for a column density 
$N_H$=2.2$\times$10$^{22}$~cm$^{-2}$). No line emission was found from 
the eastern limb; however marginal evidence for line emission in the 
halo's northern knots was found. This study illustrates the need
for deep \chan\ observations to reveal the missing SNR material
in Crab-like plerions.
\end{abstract}

\end{frontmatter}

\section{Introduction}
\label{section:intro}
\g21\ has been classified as a plerionic SNR since it has a flat radio 
spectrum,  centrally peaked radio and X-ray emission, and a highly 
polarized radio flux~\citep{green2004}.
 At a distance of $\sim$5 kpc the 80$^{\prime\prime}$-diameter
plerion would have a linear size of $\sim$2 pc.

Prior to \xmm\ and \chan\ observations there was no evidence for a 
central energy source in \g21\ and the X-ray and radio distributions 
were similar. \chan\ observations in 1999 revealed a faint X-ray halo 
($\sim$$150''$~radius) and showed that the central bright
($40''$-radius) plerion was 
surrounding a $2''$ compact centre~\citep{slane2000, safi-harb2001}.
The spectrum of \g21\ was described by a powerlaw with the photon
index increasing with distance from the central source. No evidence 
of thermal emission from the halo was found using 6 \chan\ ACIS-S 
observations~\citep{safi-harb2001}.  Early \xmm\ observations also 
showed that the halo has a nonthermal spectrum~\citep{warwick2001}.
A plerionic origin to the halo is problematic because of the larger X-ray
size when compared to the radio size, and a shock-heated supernova
material origin is also
problematic because of the absence of limb-brightening and thermal emission.
While the nature of the halo is unclear, it was suggested that 
it could be partially caused by dust scattering since G21.5-0.9 is
a  bright and heavily absorbed source~\citep{safi-harb2001}.

The X-ray halo was later modelled using \xmm\ data as a dust-scattering 
halo.  It was found that the halo could not be due to dust-scattering alone.
As well, 
a thermal component was detected in the northern knots (whose presence 
cannot be explained by the dust model), but not elsewhere in the 
halo~\citep{bandiera2004}.

\g21\ was chosen as a calibration target for the \chan\ X-ray observatory 
and, as a result there  are numerous observations publicly 
available. Here, we concentrate 
on imaging with the Advanced CCD Imaging Spectrometer (ACIS) and 
the High-Resolution Camera (HRC), as well as a spectral analysis of
the -120$^\circ$C ACIS-I data.

\vspace{-0.3cm}

\section{Imaging}
\label{section:imaging}
Images were created from observations taken between 1999 Aug.~23 
and 2004 Mar.~26 with ACIS, and between 1999 Sept.~4 and 2003 
May~17 with HRC-I.  All data processing was performed with the software 
package CIAO~2.3.  Each observation was processed individually to create 
a cleaned event file by retaining events with ASCA grades 02346 and 
rejecting hot pixels.  As well, the data was filtered for good time 
intervals, removing periods with high background rates or unstable 
aspect.  The observations were combined using the CIAO tools
$reproject\_events$ and $dmmerge$.  A summary of the exposure times for
the \chan\ observations is given in Table~\ref{table:exptime}.  The
maximum and minimum off-axis angles for the ACIS data were $9.3'$
(observation 1441) and $0.3'$ (observations 159 and 1230), respectively.  
The total integrated exposure times 
for the ACIS images and HRC-I images were 520.3~ksec (194.6~ksec with off-axis angle  $< 2'$) and 245.2~ksec (all observations had off-axis angle $< 0.4'$), 
respectively.  This is more than a magnitude deeper than previous studies.

\subsection{The Halo}
\begin{figure*}[htbp]
\begin{center}
\includegraphics[width=5.5in]
{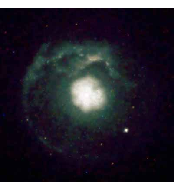}
\includegraphics[width=2.7in]{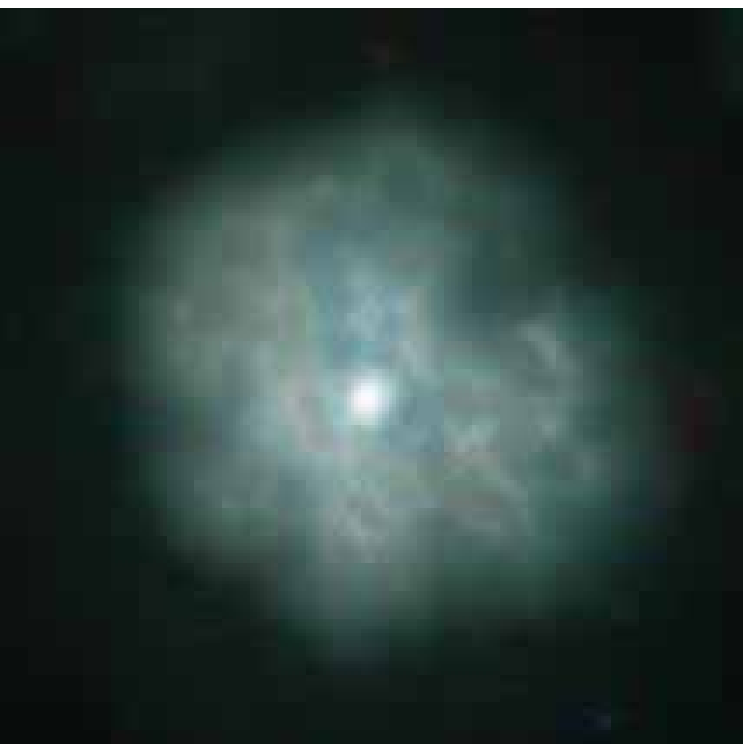}
\caption{ACIS image of \g21\ using all available ACIS data (Table~\ref{table:exptime}). Red corresponds to 0.2-1.5 keV, green to 1.5-3.0 
keV, and blue to 3.0-10.0 keV. The entire remnant is shown on the top and 
the 40$^{\prime\prime}$-radius plerion is shown on the bottom.}
\label{figure:truecolour}
\end{center}
\end{figure*}

\begin{figure*}[htbp]
\begin{center}
\includegraphics[width=2.7in]{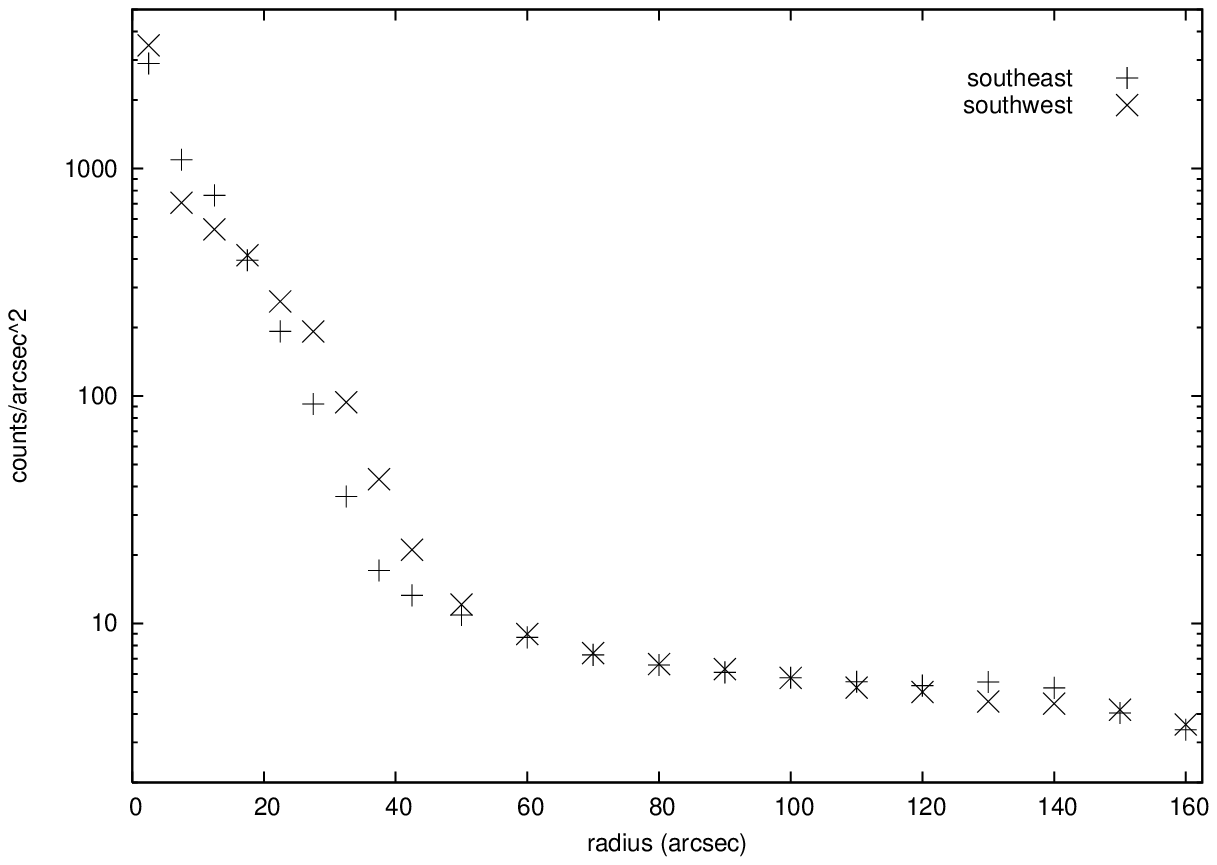}
\includegraphics[width=2.7in]{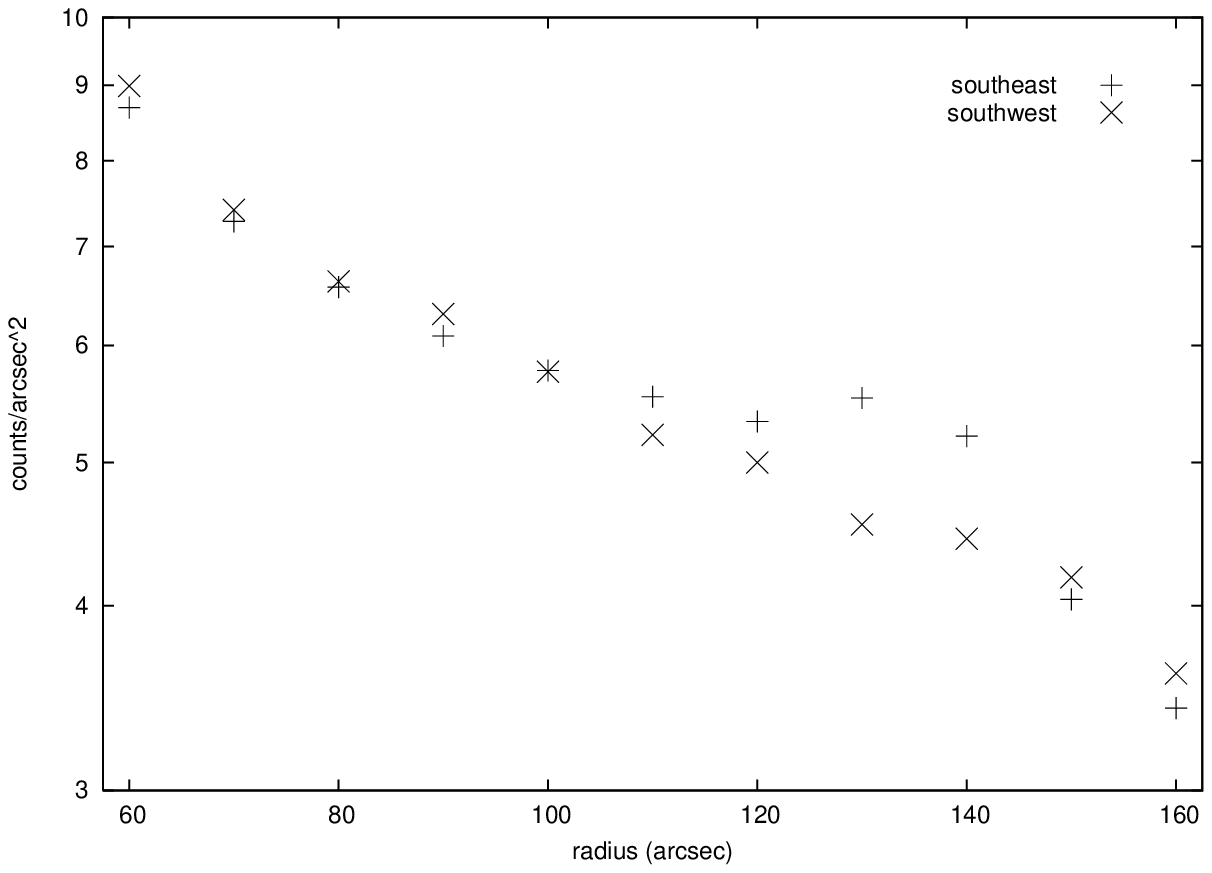}
\caption{Radial profiles of the southeast and southwest quadrants of
  \g21.  The excess emission seen in the right panel between
  $\sim$120$^{\prime\prime}$ and
  140$^{\prime\prime}$ corresponds to the limb-brightening in the east
  (see also Fig.~\ref{figure:truecolour}).}
\label{figure:radialse}
\end{center}
\end{figure*}

Using 59 ACIS observations  (269.7~ksec with ACIS-I and 250.6~ksec with 
ACIS-S), we created images of \g21\ in the soft (red, 0.2-1.5~keV), medium 
(green, 1.5-3.0~keV), and hard (blue, 3.0-10.0~keV) energy bands
(smoothed with a Gaussian distribution, $\sigma$~=~$2.5''$).  The soft energy image reveals the X-ray halo as
incomplete in the southwest, whereas the medium energy image shows a nearly 
circular halo.  The X-ray halo is least evident in the hard image. The 
images of the three energy bands were then combined to form the colour image 
in Fig.~\ref{figure:truecolour}.  Limb-brightening is observed for the first 
time at the boundary of the X-ray halo, most noticeably in the medium energy 
band and on the eastern side.
Radial profiles (Fig.~\ref{figure:radialse}) were created of the
southeast and southwest quadrants, yielding the level of limb-brightening.

\begin{figure*}[htbp]
\begin{center}
\includegraphics[width=2.7in]{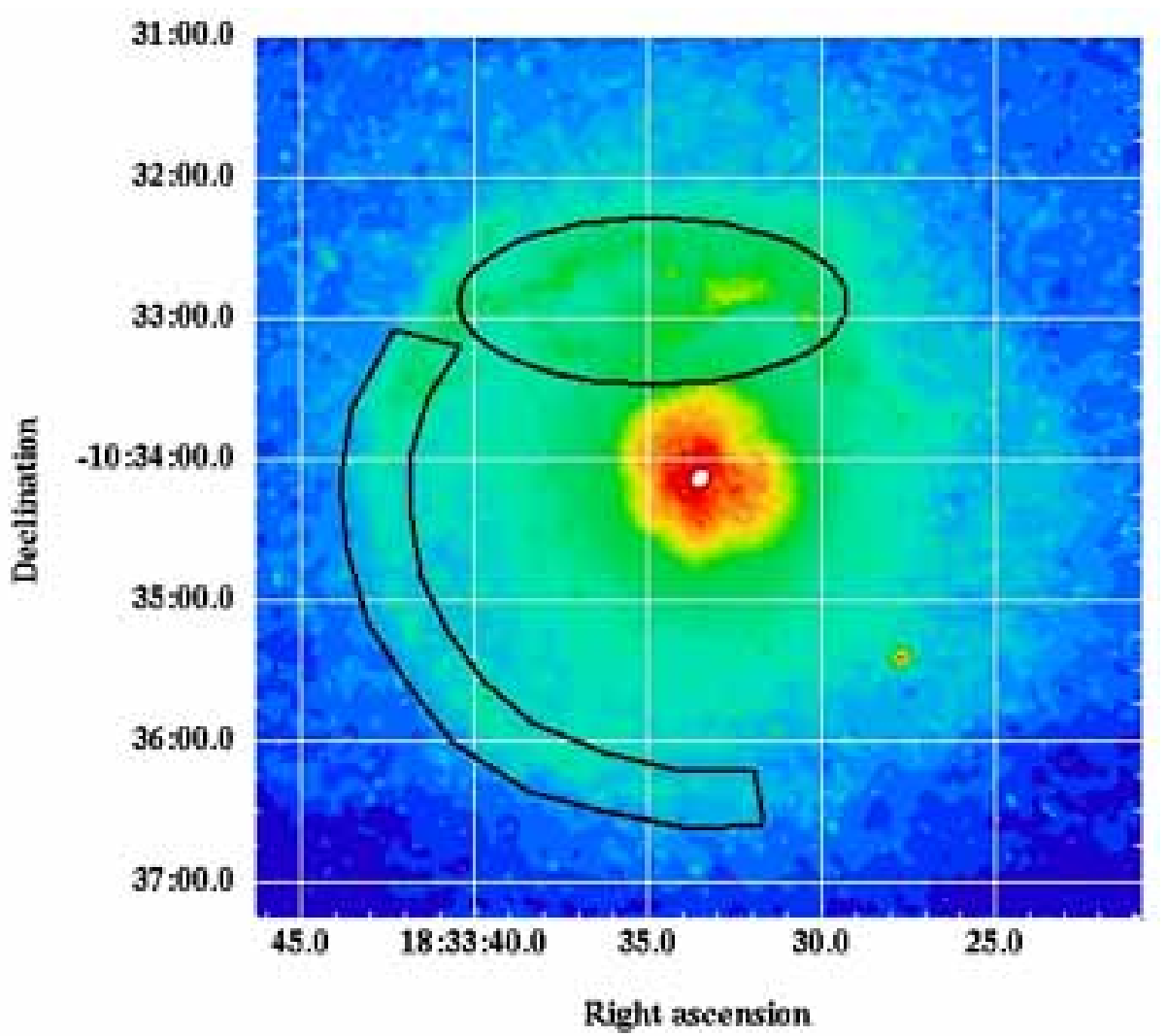}
\includegraphics[width=2.7in]{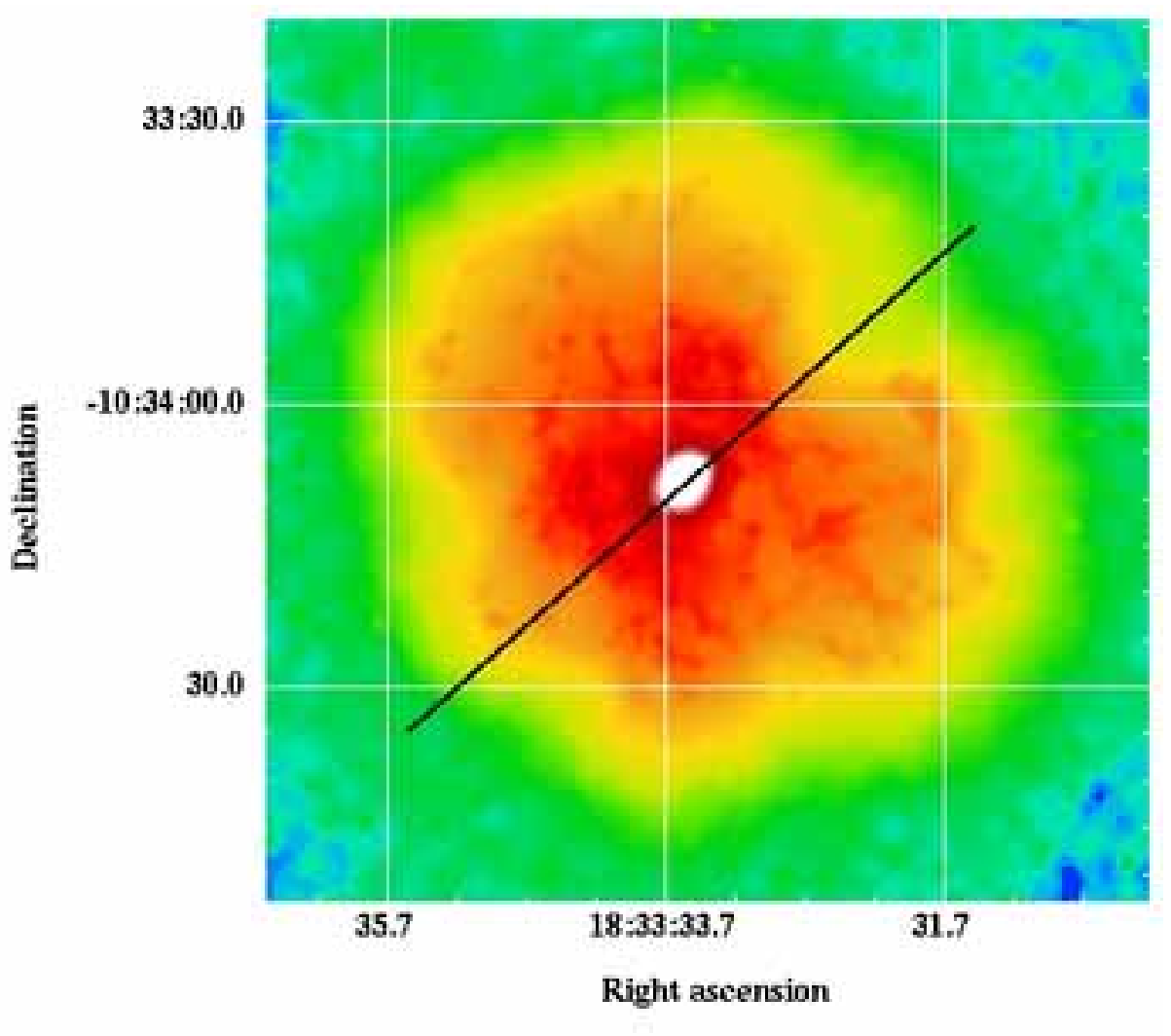}
\caption{ACIS images of \g21, revealing a faint X-ray halo 
with limb-brightening in the east.  The regions selected for spectral 
analysis were the northern knots and the eastern limb 
(see \S\ref{section:spectral_analysis}). The right image highlights the structure observed 
in the plerion.}
\label{figure:acisg21}
\end{center}
\end{figure*}

In Fig.~\ref{figure:acisg21}, we show the corresponding intensity image 
(520.3 ksec integrated exposure time). The left image spans the entire remnant 
($377''$, smoothed with $\sigma$~=~$2.5''$), while the right
image spans the central 
$94''$ (smoothed with $\sigma$~=~$1''$) and highlights the structure observed 
in the plerion.  Again, the eastern limb is visible. This shell-like
structure indicates that we have likely detected the SNR shell. 
Most notably in the halo is a knotty structure visible in the northern part 
of the halo (also referred to in previous work as the northern spur). This 
region is of particular interest as it has a radio counterpart (B. Gaensler, 
private communication). In \S3, we present its spectrum which suggests that 
it could be due to enhanced emission from swept-up supernova material.

\subsection{High-Resolution Imaging of the Inner Plerion}
\begin{figure*}[htbp]
\begin{center}
\includegraphics[width=2.7in]{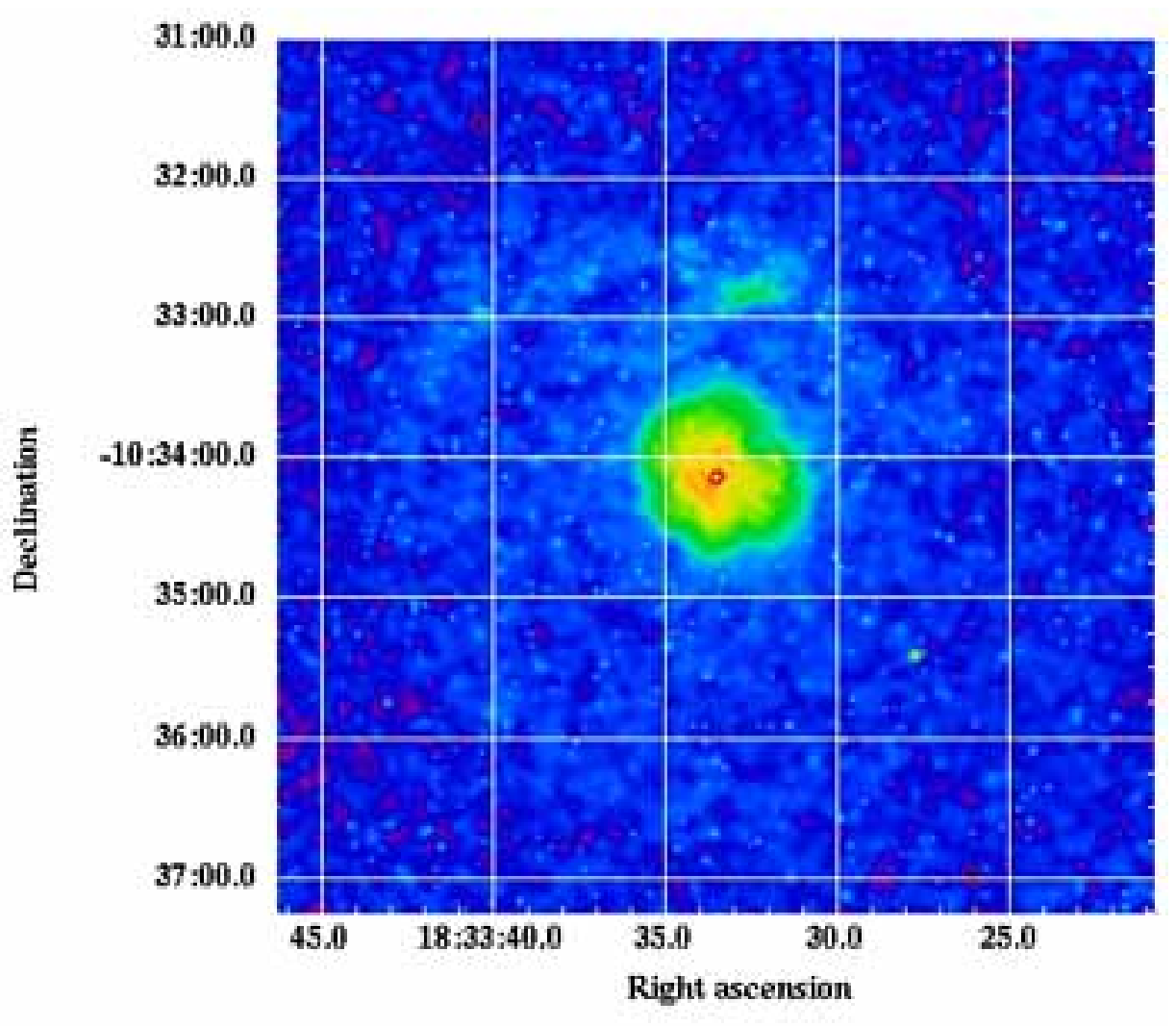}
\includegraphics[width=2.7in]{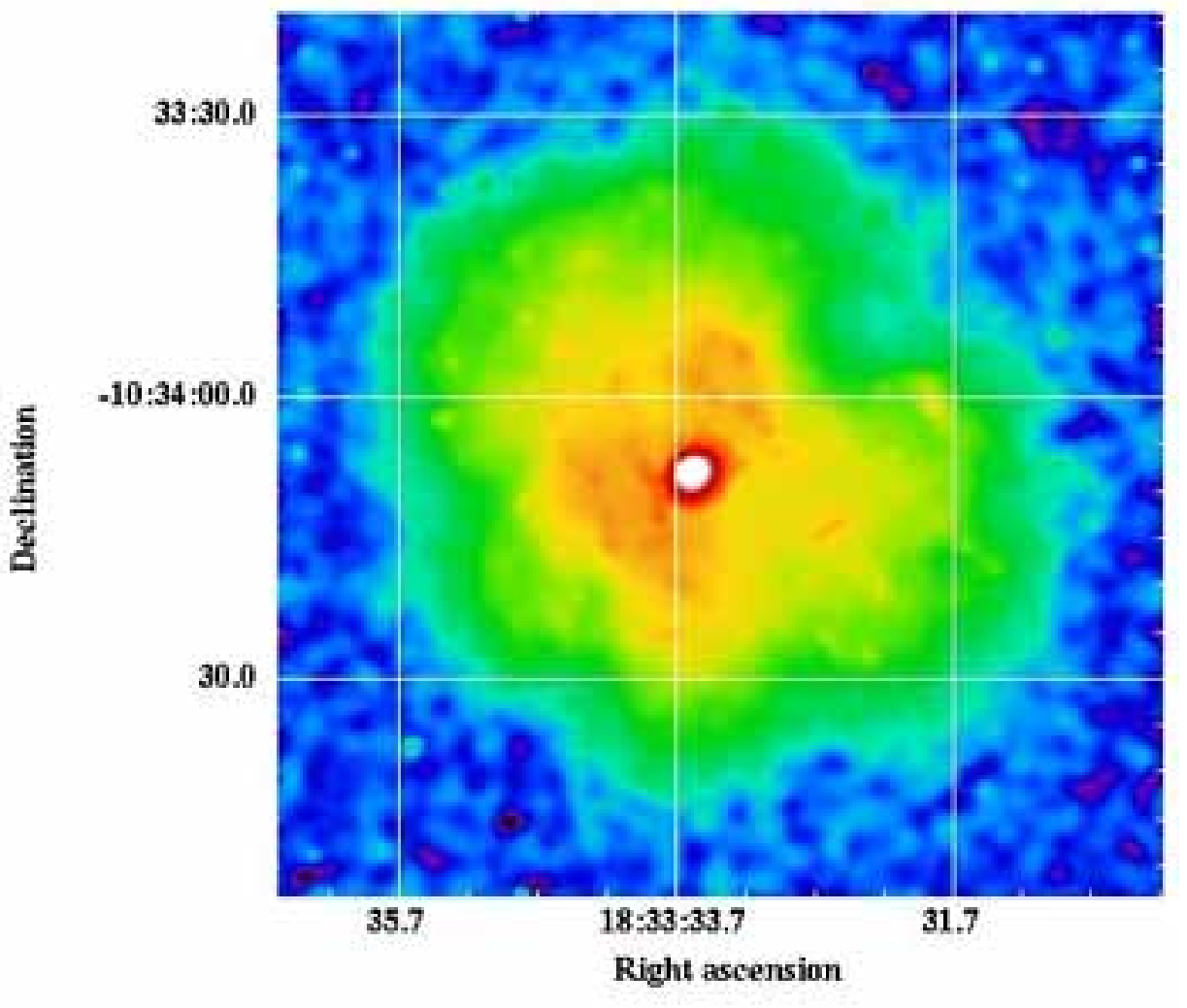}
\caption{HRC-I images of \g21.}  
\label{figure:hrcig21}
\end{center}
\end{figure*}

The bottom panel of Fig.~\ref{figure:truecolour} and the right panel of 
Fig.~\ref{figure:acisg21} show the ACIS energy color image (as in \S2.1) 
and the false-colour image of the inner bright component 
(40$^{\prime\prime}$-radius). The HRC-I image (245.2~ksec exposure, 
13 observations) is shown in  Fig.~\ref{figure:hrcig21}. 

The compact source is clearly visible at the centre of the plerion, and 
filamentary structures are noticeable features of the plerion in all 
images. In particular, the high-resolution HRC image reveals wisp-like 
features southeast of the compact source near $\alpha$=18$^h$33$^m$34.0$^s$ 
and $\delta$=-10$^\circ$34$'$10.5$''$ (J2000), and northwest of the compact 
source near $\alpha$=18$^h$33$^m$33.3$^s$ and 
$\delta$=-10$^\circ$33$'$58.0$''$ (J2000). Similar features have been
observed in known pulsar wind nebulae and are believed to result from
the deposition of 
the pulsar wind's energy into its surroundings.

Furthermore, our deep images confirm the double-lobed morphology of the 
plerion previously seen in the earlier images~\citep{safi-harb2001}. This 
structure is particularly visible in the northwest, as indicated by the line 
of symmetry running from northwest to southeast.  The indentation in the 
northwest of the X-ray emission is correlated with a trough in the radio, 
supporting the picture that the symmetric lobes could be evidence of pulsar 
jets~\citep{furst1988}. Modelling of the plerion is beyond the scope of this 
paper and will be addressed in future work. 

\section{Spectral Analysis}
\label{section:spectral_analysis}
The ACIS-I, -120$^\circ$C data (Table~\ref{table:exptime}) was filtered for 
good time intervals and ASCA  grades 02346, yielding 127.5 ksec of data 
used in the spectral analysis.  A CTI correction was performed 
on each observation using the CIAO tool $acis\_process\_events$.

The regions selected for analysis were the bright knots in the north of the 
SNR and the limb-brightening observed in the east 
(Fig.~\ref{figure:acisg21}), as well as concentric rings of increasing 
radius centred on the compact central source ($5''$ increments from 
$0''$ to~$45''$, $10''$ increments from $45''$ to~$165''$). Note that 
the bright plerion extends to $40''$, the brightest knots in the north 
of the remnant are located at $70''$-$95''$, and the limb-brightening in 
the southeast of the remnant is observed at $120''$-$150''$.

The spectra were grouped using $grppha$ to a minimum 100 counts per
bin for the northern knots and a minimum 50 counts per bin for the
eastern limb.  The 
15 spectra for a particular region were fit simultaneously using XSPEC.  
The command $group$ was used in XSPEC to produce a plot with spectra from 
all ACIS-I, -120$^\circ$C observations combined (\S3.2 and \S3.3).

\subsection{Plerion and Extended Emission}
\begin{figure*}[htbp]
\begin{center}
\includegraphics[width=1.75in]{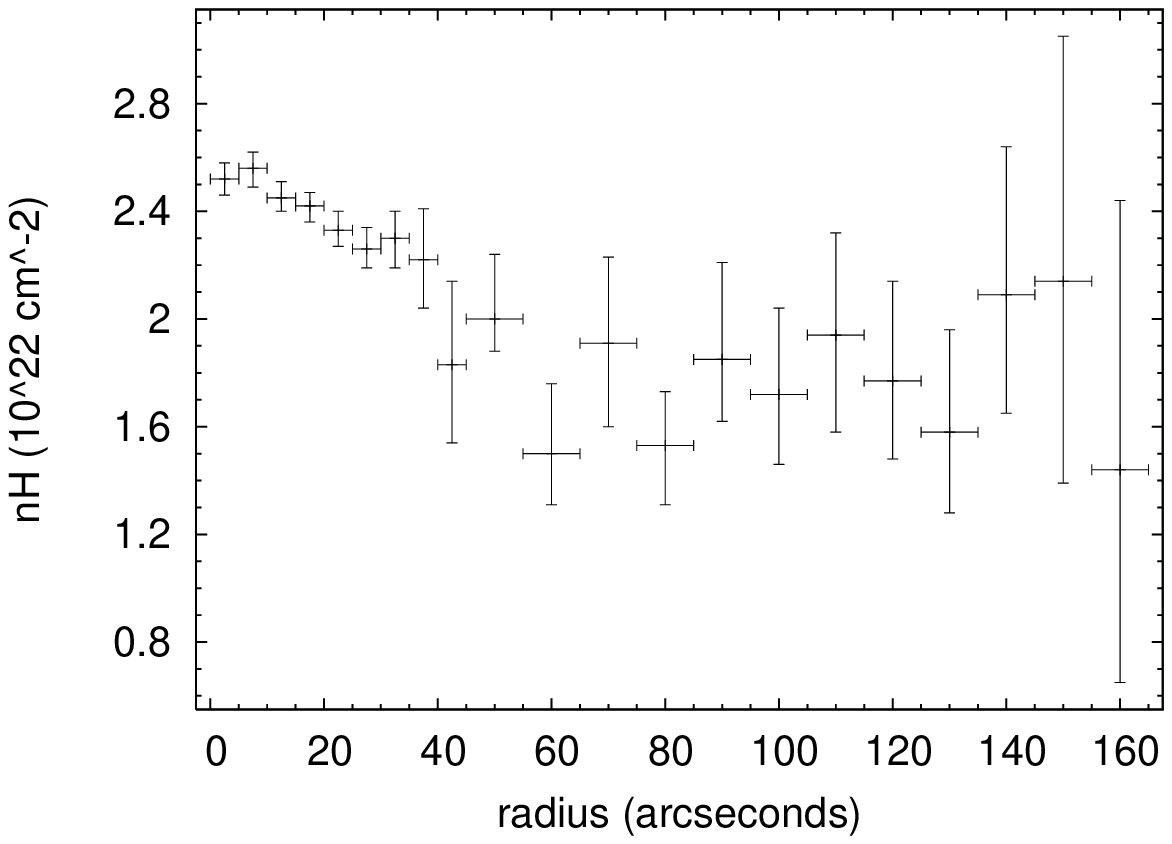}
\includegraphics[width=1.75in]{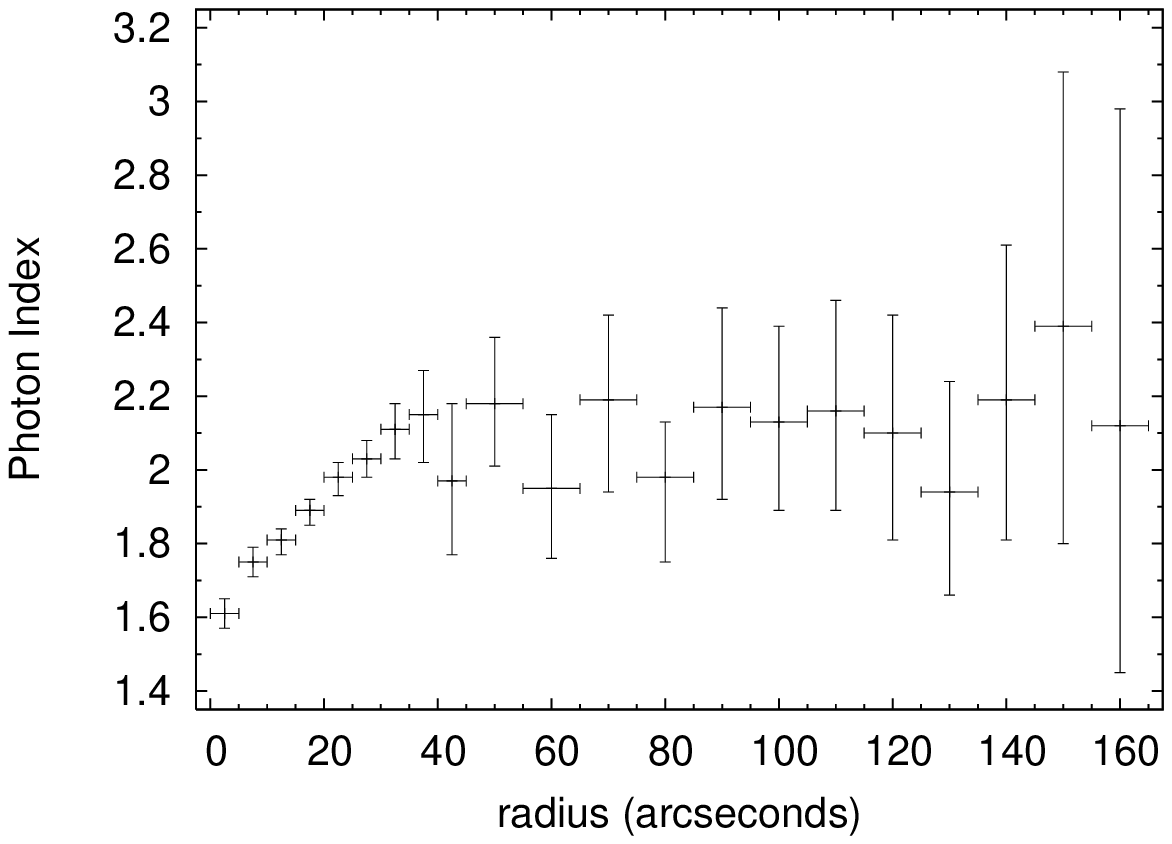}
\includegraphics[width=1.75in]{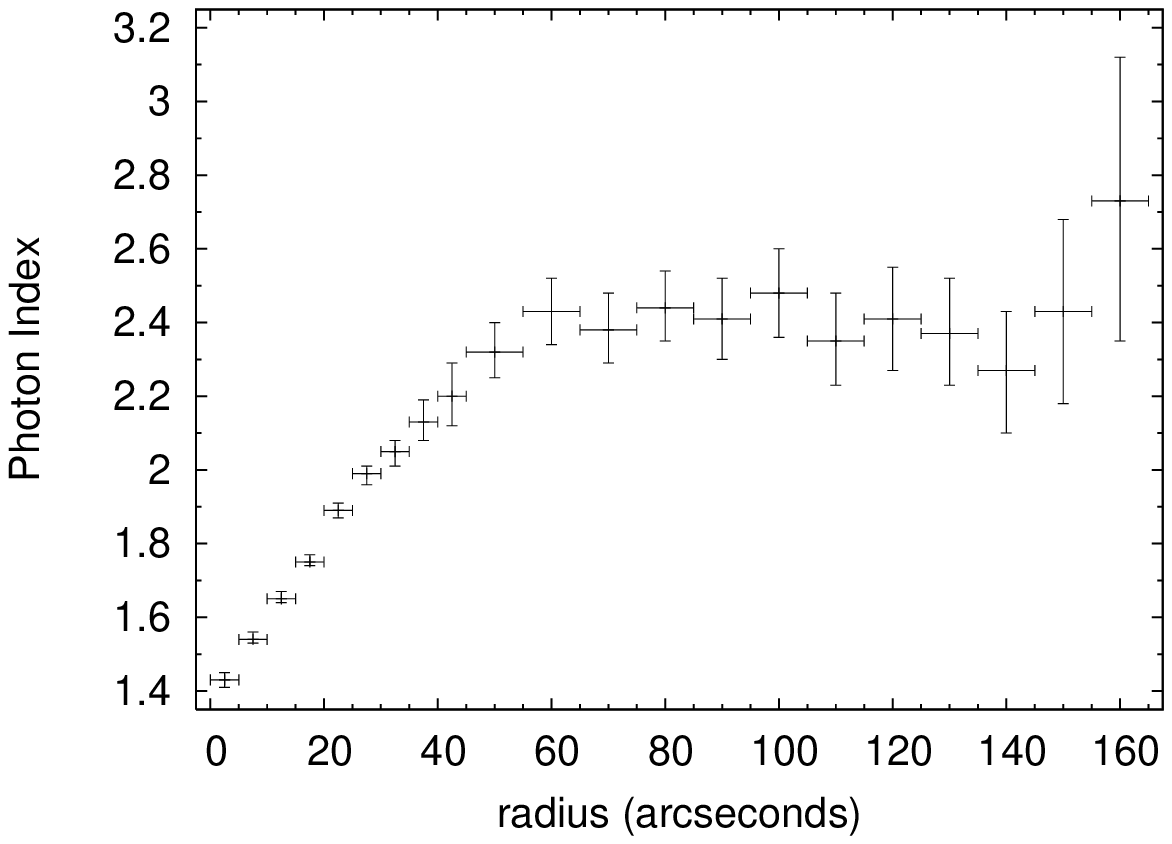}
\caption{The variation of $N_{\rm{H}}$ and $\Gamma$ across \g21. {\bf Left:} 
Radial variation of $N_{\rm{H}}$ for an absorbed powerlaw.  {\bf Centre:} 
Radial variation of $\Gamma$ when $N_{\rm{H}}$ is allowed to vary freely in an 
absorbed powerlaw.  {\bf Right:} Radial variation of $\Gamma$ when 
$N_{\rm{H}}$ is frozen at 2.2~$\times$~10$^{22}$~cm$^{-2}$ (the best
fit value) in an absorbed powerlaw model.} 
\label{figure:NHandGAMMA}
\end{center}
\end{figure*}

Spectra obtained from concentric rings, centered on the compact center at 
$\alpha = 18^h33^m33^s$ and $\delta = -10^{\circ}34'08''$ (J2000), were 
fit with a powerlaw (with $N_{\rm{H}}$ allowed to vary, and with 
$N_{\rm{H}}$~=~2.2~$\times$~10$^{22}$~cm$^{-2}$, which was the column 
density found by fitting a powerlaw to the remnant, consistent with  
~\cite{safi-harb2001}).
We found that the photon index, $\Gamma$, increases with radius from 
1.61~$\pm$~0.04 (1.43~$\pm$~0.02 with 
$N_{\rm{H}}$~=~2.2~$\times$~10$^{22}$~cm$^{-2}$) at $0''-5''$ to 
2.15~$\pm$~0.13 (2.13~$\pm$~0.06 with 
$N_{\rm{H}}$~=~2.2~$\times$~10$^{22}$~cm$^{-2}$) at the edge of the 
plerion (Fig.~\ref{figure:NHandGAMMA}, errors are 90\% confidence).  
When $N_{\rm{H}}$ is allowed to vary, the general trend is a decrease in 
$N_{\rm{H}}$ with an increase in radius (Fig.~\ref{figure:NHandGAMMA}).  
Above $40''$ the number of counts 
decreases dramatically, but $\Gamma$ seems to be constant at $\sim$2.4 
($N_{\rm{H}}$~=~2.2~$\times$~10$^{22}$~cm$^{-2}$).  
The constant value for the photon index across the halo suggests that the 
halo is unlikely to be an extension of the plerion (see \S3.2 and \S3.3).
No evidence for thermal emission was found within a radius of $40''$.

\subsection{Northern Bright Knots}

\begin{figure}[htbp]
\begin{minipage}[t]{2.6in}
\begin{center}
\rotatebox{270}{\includegraphics[width=1.7in]{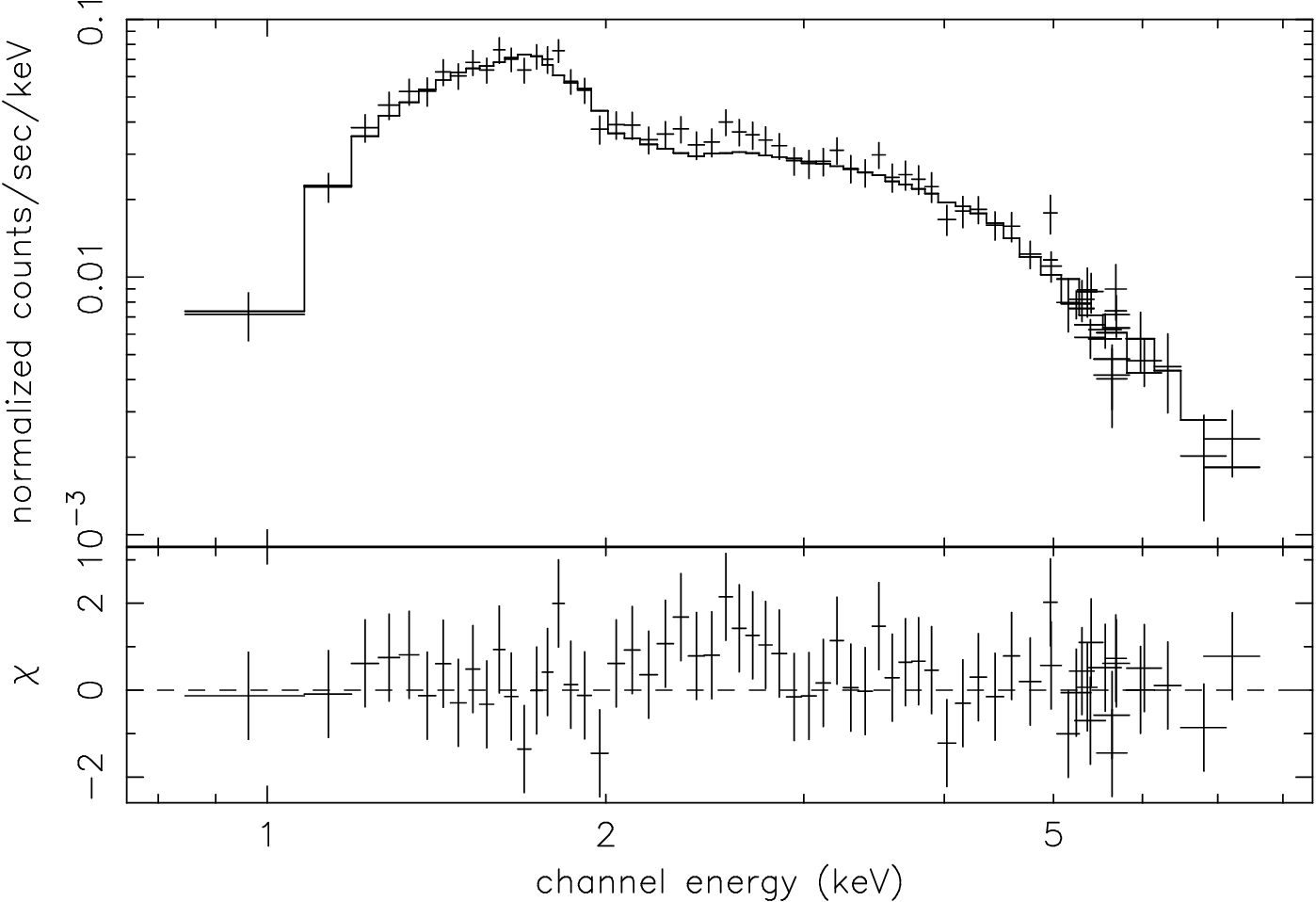}}
\caption{\label{figure:north_acisi_120_power_nHfree}Absorbed powerlaw fit to 
the ACIS-I, -120$^\circ$C data of the northern knots.  Excess emission is 
visible, indicating marginal evidence for sulfur and silicon in the knots.}
\end{center}
\end{minipage}
\hfill
\begin{minipage}[t]{2.6in}
\begin{center}
\rotatebox{270}{\includegraphics[width=1.7in]{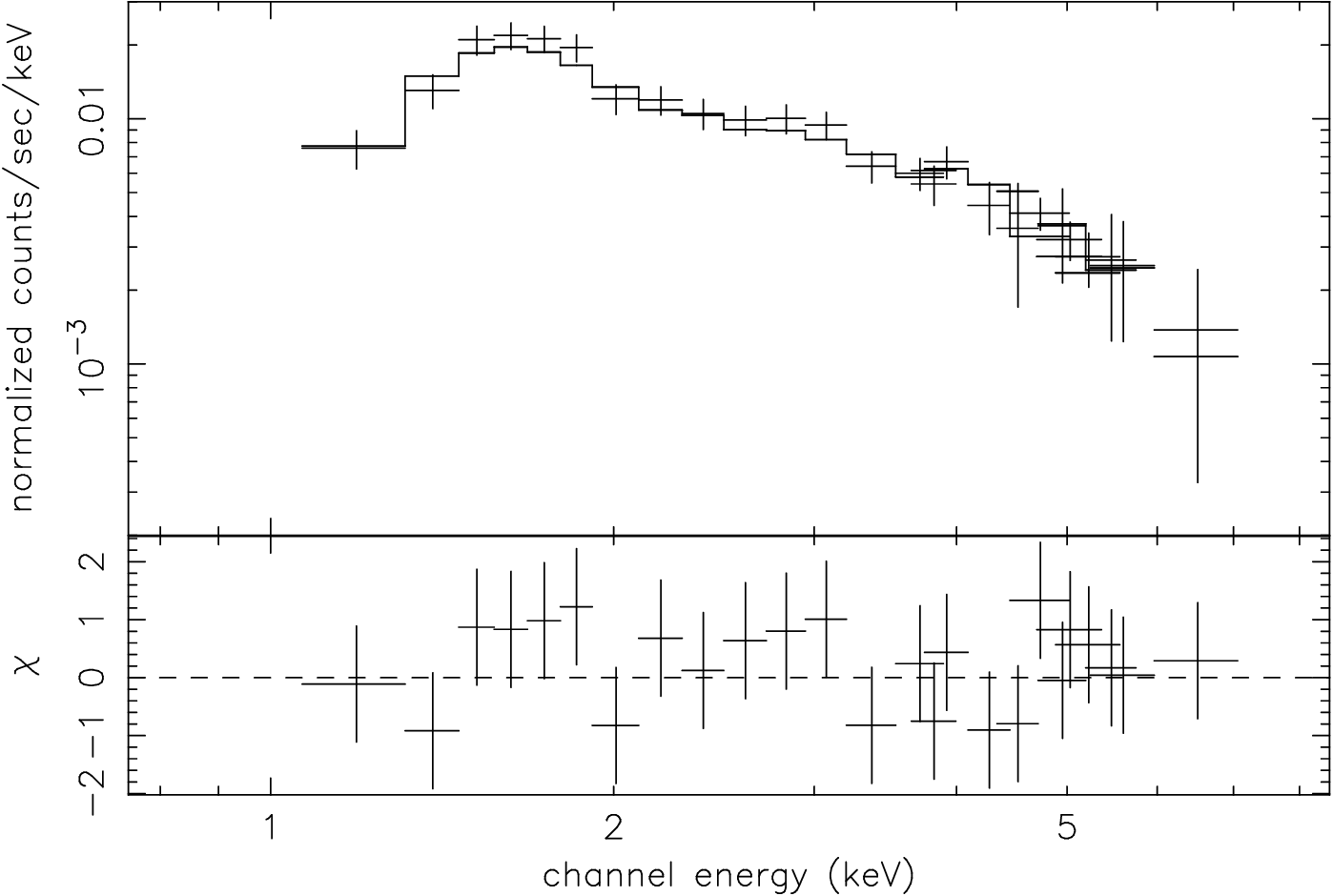}}
\caption{\label{figure:limb2_acisi_120_power_nHfree}Absorbed powerlaw fit to 
the ACIS-I, -120$^\circ$C data of the eastern outermost arc.}
\end{center}
\end{minipage}
\end{figure}

The spectrum of the bright knots in the north of the remnant 
shows marginal evidence of line emission. Fitting a powerlaw to this region 
(Fig.~\ref{figure:north_acisi_120_power_nHfree}, 
resulted in a photon index of 
$\Gamma$~=~2.07~$\pm$~0.14
($N_{\rm{H}}$~=~1.86~$\pm$~0.20~$\times$~10$^{22}$~cm$^{-2}$) with a reduced 
$\chi^2$ of 1.11 (dof=108).  Excess emission is visible near 1.84 keV, 
2.38 keV, 2.64~keV
indicating the presence of silicon and sulfur in 
the knots. Adding gaussians on top of the powerlaw at 1.84 keV, 2.38 keV 
and 2.64 keV produced F-test probabilities of 
0.845, 5.4$\times$10$^{-3}$, and 3.3$\times$10$^{-3}$
respectively, indicating that the addition of 
the gaussians significantly improves the fit. 

The XSPEC \pshock\ model was then used to search for thermal emission that 
may be due to shock-heated ejecta or interstellar matter. As shown in 
Table~\ref{table:acisi_spec}, the \pshock\ model provides an adequate 
fit; however with a lower column density, an unrealistically high 
temperature and a low ionization timescale.  Fitting a two component 
model (powerlaw+\pshock) to the bright knots, 
a good fit is also obtained with 
$N_{\rm{H}}$=2.46~(2.04-3.15)~$\times$~10$^{22}$~cm$^{-2}$ (which is
consistent with  $N_{\rm{H}}$ towards the central plerion), a photon index 
of 2.33 (2.12-2.57) and $kT$=0.14 (0.08-0.20) keV (reduced $\chi^2$=1.08, 
dof=105).
Additional \chan\ data will help constrain the thermal component parameters
(see also Bocchino, these proceedings, for an \xmm\ study of this region).

\subsection{Eastern Outermost Arc}
In Table~\ref{table:acisi_spec}, we summarize the spectral fitting results 
to the limb-brightened region in the east of the remnant; and in 
Fig.~\ref{figure:limb2_acisi_120_power_nHfree} we show the powerlaw fit.
No line emission is 
observed in this region. However, due to the low count rate, the inclusion 
of additional observations may reveal additional detail in the spectrum.

\section{Conclusions}
Deep \chan\ imaging of \g21\ reveals previously unseen detail, in both the 
plerion and the X-ray halo.  Limb-brightening is revealed in the eastern 
portion of the X-ray halo, and wisp-like structures reminiscent of shocked 
pulsar winds are observed in the plerion.  A spectral analysis performed 
with ACIS-I data shows the plerion is fit well by a powerlaw with a photon 
index that increases with radius from the centre of the plerion.  However, 
the photon index is constant across the X-ray halo, indicating that the 
halo is unlikely an extension of the plerion. Marginal evidence was found for 
line emission in the bright knots in the north of the SNR, suggesting the 
presence of swept-up supernova material.  
Future work includes a spectral analysis 
of all ACIS-S data in addition to the ACIS-I data, which could reveal 
additional detail in the spectra; a timing analysis of all 
the available observations, a correlation of the X-ray data with the 
observations at radio and infrared wavelengths, and modeling of our results.

\vspace{0.15in}
The authors acknowledge support by the Natural Sciences and Engineering
Research Council (NSERC) of Canada. H. Matheson is an NSERC PGS-M fellow 
and S. Safi-Harb is an NSERC University Faculty Award fellow. 

We thank the referees for their useful comments.

\begin{table*}[htbp]
\begin{center}
\begin{tabular}{crl} \hline
Detector & Time & Observations (off-axis angles)\\ 
\hline\hline
ACIS-I, -100$^\circ$C & 41.9 & 158 ($5.8'$), 160 ($6.7'$), 161 ($5.9'$), 162 ($1.7'$) \\ 
ACIS-I, -110$^\circ$C & 100.4 & 1233 ($1.7'$), 1441 ($9.3'$), 1442 ($5.8'$), 1443 ($5.9'$),\\
                      &       & 1772 ($7.3'$), 1773 ($5.8'$), 1774 ($4.3'$), 1775 ($2.5'$),\\
                      &       & 1776 ($0.8'$), 1777 ($5.8'$), 1778 ($4.3'$), 1779 ($2.5'$) \\ 
ACIS-I, -120$^\circ$C & 127.5 & 1551 ($2.5'$), 1552 ($2.5'$), 1719 ($7.3'$), 1720 ($5.8'$),\\
                      &       & 1721 ($4.3'$), 1722 ($2.5'$), 1723 ($0.8'$), 1724 ($5.8'$),\\
                      &       & 1725 ($4.3'$), 1726 ($2.5'$), 2872 ($3.7'$), 3473 ($7.0'$),\\
                      &       & 3692 ($2.5'$), 3699 ($3.7'$), 5165 ($3.8'$) \\
ACIS-S, -100$^\circ$C & 36.8 & 159 ($0.3'$), 165 ($6.2'$), 1230 ($0.3'$) \\ 
ACIS-S, -110$^\circ$C & 68.2 & 1433 ($1.2'$), 1434 ($6.2'$), 1769 ($1.3'$), 1770 ($1.3'$),\\
                      &      & 1771 ($1.3'$), 1780 ($4.7'$), 1781 ($4.7'$), 1782 ($4.7'$)\\
ACIS-S, -120$^\circ$C & 145.7 & 1553 ($1.3'$), 1554 ($1.2'$), 1716 ($1.3'$), 1717 ($1.3'$),\\
                      &       & 1718 ($1.3'$), 1727 ($4.7'$), 1728 ($4.6'$), 1729 ($4.6'$),\\
                      &       & 1838 ($1.2'$), 1839 ($2.4'$), 1840 ($5.3'$), 2873 ($1.2'$),\\
                      &       & 3474 ($7.0'$), 3693 ($1.2'$), 3700 ($1.2'$), 4353 ($5.2'$),\\
                      &       & 5166 ($1.2'$) \\
HRC-I & 245.2 & 142 ($0.3'$), 143 ($0.3'$), 144 ($0.3'$), 1242 ($0.4'$),\\
      &       & 1298 ($0.3'$), 1406 ($0.3'$), 1555 ($0.3'$), 1556 ($0.3'$), \\
      &       & 2867 ($0.3'$), 2874 ($0.3'$), 3694 ($0.3'$), 3701 ($0.2'$),\\
      &       & 5167 ($0.3'$) \\ \hline
\end{tabular}
\end{center}
\caption{Net exposure times (in ksec) and off-axis angles for \chan\ data.}
\label{table:exptime}
\end{table*}

\begin{table*}[htbp]
\begin{center}
\begin{tabular}{lccc} \hline
Model & Parameter & northern knots & eastern limb \\ 
\hline\hline
power & $N_{\rm{H}}$ (10$^{22}$ cm$^{-2}$) & 1.86 (1.70-2.06) & 1.86 (1.37-2.14) \\
 & $\Gamma$ & 2.07 (1.96-2.21) & 2.15 (1.81-2.52) \\
 & reduced chi-sq (dof) & 1.11 (108) & 0.993 (69) \\
\hline
power$^*$ & $\Gamma$ & 2.29 (2.24-2.34) & 2.37 (2.24-2.50) \\
 & reduced chi-sq (dof) & 1.18 (109) & 0.995 (70) \\
\hline
\pshock & $N_{\rm{H}}$ (10$^{22}$ cm$^{-2}$) & 1.56 (1.44-1.68) & 1.47 (1.11-1.83) \\
 & $kT$ (keV) & 4.98 (4.39-5.88) & 4.84 (3.39-7.28) \\
 & $n_et$ (10$^{8}$~s~cm$^{-3}$) & 1 (1-6.28) & 1 (1-15.9) \\
 & reduced chi-sq (dof) & 1.11 (107) & 1.02 (68) \\
\hline
\pshock$^*$ & $kT$ (keV) & 5.22 (4.83-5.77) & 4.90 (3.86-7.15) \\
& $n_et$ (10$^{9}$~s~cm$^{-3}$) & 9.8 (8.6-11.1) & 13.0 (8.4-21.4) \\
 & reduced chi-sq (dof) & 1.91 (108) & 1.15 (69) \\
\hline
\end{tabular}
\end{center}
\caption{Results obtained by fitting ACIS-I, -120$^\circ$C spectra 
simultaneously in XSPEC (0.5-8.0 keV, 90\% confidence ranges).  
Fig.~\ref{figure:north_acisi_120_power_nHfree} and 
\ref{figure:limb2_acisi_120_power_nHfree} show the powerlaw fits when 
$N_{\rm{H}}$ is allowed to vary. Abundances are consistent with solar 
for the \pshock\ model.  
*$N_{\rm{H}}$ frozen to 2.2$\times$10$^{22}$ cm$^{-2}$.}
\label{table:acisi_spec}
\end{table*}

\end{document}